\documentclass{aa}
\usepackage{txfonts}
\usepackage{graphicx}
\usepackage{natbib}
\usepackage{txfonts}
\usepackage{graphicx}
\usepackage{natbib}
\usepackage{epsfig}

\begin{document}
\title{Local re-acceleration and a modified thick target model of solar flare electrons}

\author{J.C. Brown\inst{1}, R. Turkmani\inst{2}, E.P. Kontar\inst{1}, A.L. MacKinnon\inst{3}, and L. Vlahos\inst{4}}

\institute{$^{1}$ Department of Physics \& Astronomy, University of
Glasgow, G12 8QQ, UK
 \email{john@astro.gla.ac.uk}\\
           $^{2}$ Department of Physics, Imperial College, London SW7 2AZ\\
           $^{3}$ Department. of Adult and Continuing Education,
            University of Glasgow G12 8QQ, UK\\
           $^{4}$ Department of Physics, University of Thessaloniki, 54006
                  Greece\\}
\date{\today}

\offprints{J.C. Brown, \email{john@astro.gla.ac.uk}}

\authorrunning{Brown {\it et al.}}
\titlerunning {Thick target model with local reacceleration}
\date{Received 20 August 2009 / Accepted 2 October 2009}

\abstract {The collisional thick target model (CTTM) of solar hard
X-ray (HXR) bursts has become an almost 'Standard Model' of flare
impulsive phase energy transport and radiation. However, it faces various
problems in the light of recent data, particularly the high electron
beam density and anisotropy it involves.} {We consider
how photon yield per electron can be increased, and hence fast
electron beam intensity requirements reduced, by local
re-acceleration of fast electrons throughout the HXR source itself, after injection.} {We show parametrically that, if net re-acceleration rates due to e.g. waves
or local current sheet electric (${\cal E}$) fields are a
significant fraction of collisional loss rates, electron lifetimes,
and hence the net radiative HXR output per electron can be substantially
increased over the CTTM values. In this local re-acceleration thick target model (LRTTM) fast electron number requirements
and anisotropy are thus reduced. One specific possible scenario
involving such re-acceleration is discussed, viz, a current sheet
cascade (CSC) in a randomly stressed magnetic loop.} {Combined MHD
and test particle simulations show that local ${\cal E}$ fields in
CSCs can efficiently accelerate electrons
in the corona and and re-accelerate them after injection into
the chromosphere.  In this HXR source scenario, rapid synchronisation and variability of impulsive footpoint emissions can still occur since primary electron
acceleration is in the high Alfv\'{e}n speed corona with fast
re-acceleration in chromospheric CSCs. It is also
consistent with the energy-dependent time-of-flight delays in HXR
features.}
 {Including electron re-acceleration in the HXR source allows an LRTTM modification of the CTTM
 in which beam density and anisotropy are much reduced, and alleviates theoretical
 problems with the CTTM, while making it more compatible
with radio and interplanetary electron numbers. The LRTTM is,
however, different in some respects such as spatial
distribution of atmospheric heating by fast electrons.}


\maketitle

\section{Basic CTTM Properties and Problems}
\label{Introduction}

 Since \citet{deJager1964} and \citet{Arnoldy1968}
the {\it Collisional Thick Target Model - CTTM}
\citep{brown1971,brown1972,hudson1972,brown1973} - of flare hard X-ray (HXR)
sources has become an almost {\it Standard Model} of flare impulsive phase
energy transport and radiation. It offers a simple and reasonably successful
description of several basic features of chromospheric HXR flares
and even some aspects of the distinct coronal HXR flares \citep{Krucker_etal2008}.
These include prediction/explanation of: footpoint sources; decreasing HXR source
height \citep[e.g.][]{aschwanden2002,brown2002} and source area
\citep{Kontar_etal08} with increasing energy; electron time-of-flight
energy-dependent delays in HXR light curves \citep{Aschwanden2004}.

However, a number of papers (e.g. \citet{Brown_etal90}) have reviewed problematic aspects of the standard CTTM model and aspects of recent data [especially from RHESSI - \citet{lin2002}]
certainly require modification of the most basic CTTM involving a single
monolithic loop. These include: the motion of HXR footpoints
\citep{Fletcher_etal2004}; the smallness of the albedo
component in HXR spectra \citep{KontarBrown06mirror,Kasparova_etal2007} compared to that
expected from the strong downward beaming in the CTTM \citep{brown1972};
the relative time evolution of the heated soft X-ray (SXR) plasma
emission measure $EM(t)$ and temperature $T(t)$, \citep[e.g.][]{Horan1971,Stoiser_etal2008}.
In addition the difference
between interplanetary and HXR source electron spectral indices  is
inconsistent with the CTTM prediction \citep{Krucker_etal2007,Krucker_etal2009}.
These suggest the need for more complex models involving e.g.  dynamic
filamented structures, rather than static monolithic ones. and non-collisional
effects in electron transport. In terms of theory the main CTTM problems
are the large fractional instantaneous density of the electron beam in the corona and the
time integrated total number of electrons injected \citep[e.g.][]{BrownMelrose1977,BenkaHolman1994,BenzHilaire2003}.
The beam density problem has worsened as estimates of
the beam (HXR footpoint) area have decreased \citep[e.g.][]{FletcherWarren2003},
though the \citet{Kontar_etal08}
finding that the HXR source area increases rapidly with height may
alleviate this. These problems arise from three factors \citep[cf][]{MacKinnon2006}:
(a) The high beam intensity demanded by the inefficiency of collisional
bremsstrahlung compared with long range
Coulomb collisional heating of the plasma. This problem is worsened
\citep{MacKinnonBrown1989} when additional energy {\it loss} processes are included such as
return current dissipation \citep[e.g.][]{Emslie1981,ZharkovaGordovskyy2005},
Langmuir wave generation \citep[e.g.][]{HamiltonPetrosian1987,MelNik_etal1999,Kontar2001,KontarReid2009},
masering \citep[e.g.][]{MelroseDulk1982,MacKinnon_etal1992}, electron-whistler interaction
\citep[e.g.][]{StepanovTsap2002}, Weibel instability \citep[e.g.][]{Karlick2009}, etc;
(b) the TTM injection assumption that no acceleration occurs in the radiation region so
that each injected electron radiates only once and for a time no longer than
 its collisional lifetime $t_{coll}$; (c) injection of the intense
beam is assumed to occur from a tenuous coronal accelerator.

The term CTTM is in fact used in two different ways in the flare
literature. Physically a (collisional) thick target is simply one in
which the radiating electrons lose all their energy (collisonally)
irrespective of geometry \citep{brown1971}. However, the term is often
used with reference to a particular geometry \citep{brown1972,brown1973,hudson1972}
where electrons are injected downward into the dense chromospheric
target after acceleration in the tenuous corona. Here we will mainly
address this geometry though our basic considerations of collisional
and non-collisional transport are relevant to at least some
of the types of coronal HXR source reviewed
by \citet{Krucker_etal2008}. Our discussion also applies both to HXR footpoints of static
monolithic loops and to the scenario described by \citet{Fletcher_etal2004},
where footpoint HXR sources move, probably as a
result of progressive magnetic field line reconnection.

In this paper we discuss problems (a) - (c) above and propose a
modified thick target scenario involving similar geometry and
injection but replacing assumption (b) by a re-acceleration
process acting within the HXR radiating volume. The
resulting increase in electron lifetime to much greater than $t_{coll}$
increases the photon yield per electron and reduces the necessary
electron replenishment rate, beam density, and anisotropy.
\section{HXR Source Requirements}
\subsection{Model-independent Nonthermal Emission Measure}
The instantaneous bremsstrahlung output $J(\epsilon)$ (photons s$^{-1}$ per
unit photon energy  $\epsilon$) from a source volume $V$, with local
plasma density $n_p(\bf r)$, and fast electron flux spectrum $F(E,{\bf
r})$ at position $\bf r$ and bremsstrahlung cross section
$Q_B(\epsilon,E)$ differential in $\epsilon$ is (Brown 1971)

\begin{equation}
\label{Jdef} J(\epsilon) = \overline n V \int_\epsilon^\infty
\overline F(E)Q_B(\epsilon,E)dE
\end{equation}
where the source means are $\overline n=\int_Vn({\bf r})dV/V;
\overline F(E)= \int_V n({\bf r})F(E,{\bf r})dV/(\overline nV)$. For
prescribed $Q_B$, $J(\epsilon)$ is thus related uniquely to the quantity
$\overline n V\overline F(E)$ regardless of how $\overline
F(E)$ is produced \citep{Brown_etal2003}. The spectral shape of $J(\epsilon)$
is fixed by the shape of $\overline F(E)$ while the absolute scale
of $J$ is fixed by a spectrum dependent factor of order unity times
$\overline n V \overline F_1$ where $\overline F_1=\int_{E_1}^\infty
\overline F(E)$ is the total mean electron flux above some
reference energy $E=E_1$. Following \citet{Brown_etal2003},
$\overline n V\overline F(E)$ is often used as the fundamental
unknown 'source' function in inference of HXR electron spectra from
data on $J(\epsilon)$ \citep[e.g.][]{Piana_etal2003,Kontar_etal2004,Massone_etal2004,Brown_etal2006}.
An equivalent HXR source property
which is more readily envisaged physically than the total $\overline
n V \overline F_1$ is the total {\it nonthermal emission measure} of
electrons of $E\geq E_1$ given by

\begin{equation}
\label{EM1def}
 EM_1=\int_V n_1({\bf r})n({\bf r})dV= {\overline{nn_1}V}\simeq \overline n_1
\overline n V
\end{equation}
where $n_1({\bf r})=\int_{E_1}^\infty \overline F(E,{\bf r})dE/v(E)$
is the local density of electrons of $E\geq E_1$. $EM_1$ can
readily be used for example to find the mean fractional density $f_1$ of fast electrons if the total (thermal) emission measure $EM
\simeq \overline n^2V$ in $V$ is known, viz $\overline
n_1/\overline n=EM_1/EM$. Note also that we can write
$EM_1=\overline n{\cal N}_1$ where ${\cal N}_1$ is of order the
total number of fast electrons in $V$. Numerically, in a typical
large event, the necessary $EM_1$ is $ > 10^{46}$ cm$^{-3}$ for
$E_1=20$ keV so any model of an intense HXR source must involve
conditions satisfying

\begin{equation}
\label{EM1val} EM_1 = {\overline nn_1}V=f_1 {\overline {n^2}}V =
10^{47}f_1n_{10}^2V_{27} > 10^{46} {\rm cm}^{-3}
\end{equation}
with $n=10^{10}n_{10}$ cm$^{-3}$, $V=10^{27}V_{27}V$ cm$^3$ etc.
This shows that coronal sources alone can only generate large HXR
bursts if they have unusually large volume and/or density
\citep[e.g.][]{VeronigBrown2004,Krucker_etal2008}. Maintenance of this $EM_1$ in the CTTM case requires that electrons be injected at a rate [cf Equation (\ref{Jbasic}) below and \citep{BrownEmslie1988}] ${\cal F}_1 > 10^{36}$
s$^{-1}$ above 20 keV. This large value is the origin of : (i) the
problematically large number of total electrons processed by the
accelerator during event duration $\tau_o$, viz. $\simeq {\cal
F}_1\tau_o$ or around $10^{39}$ in a few 100 s, equal to $100
\times$ the total electrons in a loop of $V=10^{27}$cm${^3}$,
$\overline n =10^{10}$ cm${^{-3}}$; (ii) the high beam density
$n_1={\cal F}_1/Av_1$ over area $A$. For $A=2\times 10^{16}$ cm$^2$
(or $\simeq$ 2 $\arcsec$ square) this gives $\overline n_1 \simeq
10^{10}$cm$^{-3}$, a density as high as the coronal loop plasma density
in which the intense CTTM beam propagates.

\subsection{Electron Lifetime and Model-dependent Replenishment Rate}
The instantaneous values of $\overline n V \overline F(E), ~ EM_1$ etc
in practice change as the electrons evolve. In cases where
electron lifetimes $\tau$ are short (compared to event duration or
observational integration times) it is necessary to sustain
$\overline n V \overline F(E), EM_1$ etc and hence $J$, by
maintaining the numbers of the electrons of $E\geq E_1$ at a rate given
roughly by

\begin{equation}
\label{rate} {\cal F}_1\simeq {\cal N}_1/\tau=EM_1/(n\tau)
\end{equation}

This can be either by injection of fresh electrons
from outside the HXR source to replace decaying ones (as in tbe
CTTM), or by a local reaccelation process acting on those inside the
source to offset their energy losses. The latter option has received very little attention in the HXR source
literature and is the one we focus on in this paper. In the case of
the CTTM model, maintenance of ${\cal N}_1$ is by replenishing injection from the
corona and $\tau=\tau_{CTTM}$ here is the electron collision time

\begin{equation}
\label{taucoll} t_
{coll}(E_1)=\frac{2}{n_{10}}\left(\frac{E_1}{20 {\rm keV}}\right)^{3/2} s=\frac{0.002}{n_{13}}\left(\frac{E_1}{20 {\rm keV}}\right)^{3/2}
\end{equation}

Since $t_{coll} \propto 1/n$, by Equation (\ref{rate}) the
injection rate ${\cal F}_1$ required to sustain $EM_1, \overline
F(E)$ is independent of $n$ \citep{brown1971}. If $\tau$ is reduced
below $t_{coll}$ by non-collisonal losses then the necessary
${\cal F}_1$ is increased and the problems of the CTTM worsened. Of
much greater interest are situations where the lifetime $\tau$
inside the HXR source is somehow enhanced over $t_{coll}$ because
one can then (Equation (\ref{rate})) attain the same $EM_1$ for a smaller replenishment rate ${\cal F}_1$ but the same instantaneous total ${\cal N}_1$.

While increasing $\tau$ reduces ${\cal F}_1$ the
consequences for fast electron density $\overline n_1$ in the source
depend on the geometry of their propagation. For example, if the
fast electrons were being injected into a HXR source from above,
increasing $\tau$ while containing them
in the same $V,n$ (eg by scattering or magnetic trapping),would leave the fast electron number density unchanged
but they would last longer and sustain  $EM_1$ for smaller ${\cal
F}_1$. If, on the other hand, they propagated freely downward, their
longer $\tau$ would cause them to penetrate more deeply, increasing
the HXR source $V$ and sustaining $EM_1$ with a smaller $n_1$ but
larger $V$. We discuss the latter situation again in Section 4.

The above estimates of the necessary ${\cal F}_1$ etc in terms of a
single $\tau$ value are only approximate. To get a more accurate
picture of how electron supply requirements are modified by
non-collisional energy losses and gains it is necessary to look more
closely at actual photon yield and its relation to electron
trajectories $E(t)$.

\subsection{Electron Trajectories and Photon Yield}

In general the number $\zeta(\epsilon)$ of photons per unit
$\epsilon$ emitted during the lifetime of an electron of initial
energy $E_*$ is

\begin{equation}
\label{zetadef} \zeta(\epsilon,E_*)=\int_{t(E\ge\epsilon)}n({\bf
r}(t))v(t)Q_B(\epsilon,E)dt
\end{equation}
where $n({\bf r}(t))$ is the plasma density along the electron path,
and $v(t)=(2E(t)/m_e)^{1/2}$ the electron speed while
$t(E\ge\epsilon)$ is the total of all intervals during which
$E(t)\ge\epsilon$. As electrons tend to decay to $E < \epsilon$
(or escape) they have to be maintained at a spectral rate ${\cal
F}_*(E_*)$ (s$^{-1}$ per unit $E_*$) to sustain the value of $\overline n V
\overline F(E)$ and hence $J(\epsilon)$. $J(\epsilon)$, ${\cal
F}_*(E_*)$ and ${\overline n} V {\overline F}(E)$ are inter-related
by \citet{brown1971,BrownEmslie1988}

\begin{eqnarray}
\label{Jbasic} J(\epsilon)=& \int_\epsilon^\infty{\cal
F}_*(E_*)\zeta(\epsilon,E_*)dE_* & = \nonumber\\ & \overline nV
\int_\epsilon^\infty \overline F(E)Q_B(\epsilon,E)dE &
\end{eqnarray}
where $\zeta(\epsilon, E_*)$ is now the {\it mean} value for a
large number of electrons of the same initial $E=E_*$ since in general
$E(t)$ can differ greatly between electrons of the same $E_*$ especially
in the case of stochastic acceleration - see below.
[Even for purely Coulomb collisional transport there is dispersion
in $E(t)$ for given $E_*$ due to the spread in
impact parameters and the finite thermal speed of target particles. Both
of these are small in the CTTM and are usually neglected - \citep[e.g.][]{brown1971}]
Clearly the ${\cal F}_*(E_*)$ necessary for given $J(\epsilon)$ is
related to $1/\zeta$ or, crudely, to $1/\tau$ as discussed in
Section 2.2. Note that when ${\cal F}_*$ varies on timescales shorter
than the electron time of flight in the HXR source, Equation
(\ref{Jbasic}) has to be modified to allow for energy dependent time
delays between features in ${\cal F}_*$ and in $J$ -i.e.
acceleration and propagation effects are convoluted in time. This
has been discussed in the collisional case by \citet{Emslie1983,Aschwanden2004}.
It is even more relevant to the
situations discussed here which specifically involve extended
electron lifetimes $\tau$.

In the CTTM, with radiation only in the collisional propagation
region (and no acceleration), the mean $dE/dt=\dot E=\dot E_{coll}= -Knv/E$ where
$K=2\pi e^4\Lambda$ with $\Lambda$ the Coulomb logarithm. The mean $E(t)$ is
thus monotonic so the maximum $E=E_*$ is the initial/injection energy and we can
write $dt=dE/(-{\dot E})=dE/|\dot E|$ and replace the $t$ integration
(\ref{zetadef}) by the $E$ integration

\begin{equation}
\label{zetaCTTM} \zeta_{CTTM}(\epsilon,E_*)=
\frac{1}{K}\int_\epsilon^{E_*}EQ_B(\epsilon,E)dE
\end{equation}

It is the small value of $\epsilon Q_B(\epsilon,E)$ compared with
$K/E^2$ here that makes collisional bremsstrahlung an inefficient
source of HXRs in any model, and demands large electron injection
rates and beam power. Even if we can increase the electron lifetime
and reduce the necessary number supply rate {\it the power required
is unchanged} or may even be increased.

Any non-collisional transport process which acts solely to add
energy {\it losses} $\dot E$ to the collisional ones can only reduce
$\zeta$ below $\zeta_{CTTM}$ and so increase the necessary ${\cal
F}_*$ and power requirements \citet{MacKinnon2006}. The only processes capable of allowing
$\zeta>\zeta_{CTTM}$, hence reducing ${\cal F}_*$, are ones which
tend on average to increase the mean electron lifetimes over
$t_{coll}$. (We see below that the actual effect of this on $\zeta(\epsilon)$ depends
on the form of $E(t)$ and of $Q_B(\epsilon,E)$). Physically this
corresponds to acceleration inside the HXR source, a process rather
arbitrarily excluded in conventional CTTM assumptions. The effect on
$\zeta$ of changing $\dot E$ is not immediately obvious as we show by considering some
simple parametric forms $\phi(E)$ to describe the effect of the
acceleration relative to collisions, viz.

\begin{equation}
\label{dotEgen}\dot E = \dot E_{noncoll}+\dot E_{coll}=\phi(E)\dot
E_{coll}
\end{equation}

To measure the effect of varying $\phi$ on $\zeta$ we have to adopt
a specific form for $Q_B(\epsilon,E)$ and we first consider the
Kramers form $Q_{BK}=Q_o/\epsilon E$ ($\epsilon \le E$ ) with $Q_o$ a
constant \citep{Kramers1923},
for which a measure of $\zeta$ is the quantity

\begin{equation}
\label{xidef}
\xi=\frac{K}{Q_o}\zeta=\int_\epsilon^{E_{max}}\frac{dE}{\phi(E)}
\end{equation}
This simplification lets us give several illustrative analytic
examples of the dependence of $\zeta$ on trajectories $E(t)$
\citep[cf][]{BrownMacKinnon1985}. The true $Q_B$ behaves in a more complex way
the consequences of which we mention below. For collisions only
(CTTM), $\phi=1$ and $\xi = E_{max}-\epsilon$. Other informative
cases are -

\begin{itemize}
\item (i) $\phi(E)=0 ~~ \forall E \Rightarrow \xi \rightarrow \infty$
since the electron formally has infinite lifetime. Physically this contrived
idealisation would be like dragging an electron at constant speed through the
plasma, energy supply exactly offsetting losses and making $\tau\rightarrow
\infty$
\item (ii) $\phi(E)=$ constant $C$

(a) $C>0$ (net energy loss) $\Rightarrow \xi =
\frac{E_o-\epsilon}{C}=\frac{\xi_{coll}}{C}$ so that $\zeta$ is
only enhanced in this case for $0<C<1$ which is also unrealistic corresponding to to fine tuning of $\dot
E_a$ to partially offsett losses $\dot E_{coll}$ but not reverse
them to a net gain.

(b) $C<0$ (net energy gain). Here $E(t)$ increases indefinitely
($E_{max} \rightarrow \infty $), as $t\rightarrow\infty$ and $\xi
=(E_{max}-\epsilon)/|C|\rightarrow \infty$. Though an infinite
lifetime is clearly unphysical, arbitrarily increased $\zeta $ is
possible if arbitrarily high $E_{max}$ is reached.

\item (iii)  $\phi(E) = -\phi_1(E/E_1)^a$ with $\phi_1>0$ (net energy gain). Here again
there is a formally infinite lifetime with  $E_{max}\rightarrow \infty$ as
$t\rightarrow \infty$ but, for $a\neq 1$,
\begin{equation}
\label{xival}
\xi=\frac{E_1}{(a-1)\phi_1}\left|\left(\frac{\epsilon}{E_1}\right)^{-a+1}-\left(\frac{E_{max}}{E_1}\right)^{-a+1}\right|
\end{equation}
This diverges for $a\le 1$ but is finite $\forall a>1$ despite the
infinite lifetime ($E_{max}\rightarrow\infty$). This is because
$Q_{BK} \propto 1/E$, with maximum value at $E=\epsilon$ so that the
contribution to $\xi$ falls as $E$ increases and the total is finite
for any sufficiently fast acceleration ($a>1$).
\end{itemize}

These examples show how $\zeta$ can depend on the specific form of the electron
trajectory $E(t)$. In addition, $\zeta$ depends on the form of $Q_B$
in relation to $E(t)$. For any $\epsilon$, $\zeta$
will be largest when $E(t)$ maximises the time spent near the value
of $E$ where $Q_B$ peaks. For the Kramers $Q_{BK}$ used above this
is at $E=\epsilon$ but even for the next simplest approximation -
the non-relativistic Bethe Heitler form $Q_{BBH}$ - the peak is
substantially shifted to $E\approx 1.7\epsilon$ as is also the case
for the full cross section as given by \citet{Haug1997} - see Figure 1.
Different forms of $E(t)$ convolved with these $Q_B$ can result in
substantial differences in rates of emission and hence in $\zeta$.

\begin{figure}
  \resizebox{\hsize}{!}{\includegraphics{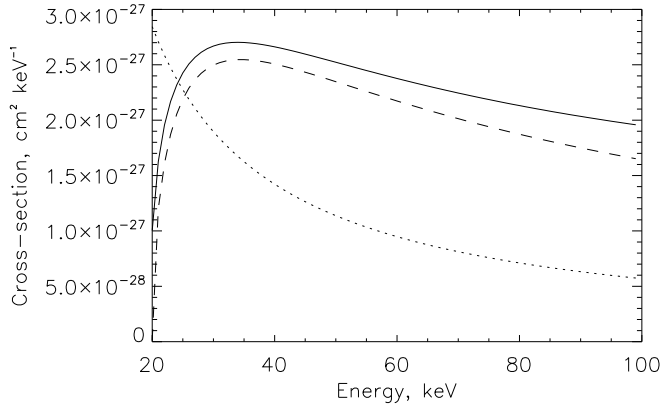}}  \caption{Cross section $Q_B(\epsilon,E)$ at $\epsilon=20$ keV showing how
  the emission contribution per unit $\epsilon$ peaks at different $E$ for
  different $Q_B$. (Dotted = Kramers, dashed = Bethe Heitler, solid = exact \citep{Haug1997})}
 \label{f1}
\end{figure}
These special cases illustrate how $\zeta$ can be enhanced by
re-acceleration $\dot E_a$ either by making the net loss rate $|\dot E| < |\dot
E_{coll}|$ or by creating a net gain rate so that $E_{max}>E_*$, both increasing the electron lifetime at $E>\epsilon$.
This reduces the necessary injected beam density and total numbers of
electrons by prolonged re-acceleration of them in the HXR
source after injection. We call this the Local re-acceleration Thick
Target Model (LRTTM). In Section 3 we discuss a specific physical
energy release scenario where these LRTTM requirements may be met.
We emphasize again that reducing ${\cal F}_1$ in this way does not
reduce the power that has to be delivered. This is always at least
the value in the CTTM since, for every erg of collisional
bremsstrahlung emitted, of order $10^5$ erg go into long range
collisional energy losses. However, in the LRTTM, most of that power
is delivered in the HXR source rather than in an external
accelerator/injector of electrons as in the CTTM case, a point to
which we return in Section 4.

To see whether and to what extent this happens in any particular LR
scenario we have to recognise that the actual photon yield
$\zeta(\epsilon)$ during the lifetime of an electron in such
scenarios is more complicated than discussed above. The $t$ integral
in equation (\ref{zetadef}) cannot be written simply as an integral
over $E$, as it can in these cases, since:
\begin{enumerate}
\item $E(t)$ is no longer monotonic in general, with $\dot E$ taking
values $>0$, $<0$, or $0$ at different parts of its path. Then the
$t$ integral can only be written as a sum of $E$ integrals with one
for each $t$ segment in which $E(t)$ is monotonic (with $\dot E
>0$ or $\dot E<0$) plus integrals over $t$ itself when $\dot E=0$ so
that $dt$ does not transform to a finite $dE$. In practice one
reverts to the basic $t$ integration (\ref{zetadef}).
\item Even if the change in variable from $t$ to $E$ is useful, the upper limit in the $E$ integrals is no longer the initial
energy $E_*$ (as it is in the CTTM) but the maximum value
$E_{max}(E_*)$ reached during the electron lifetime at $E\ge
\epsilon$
\item In re-acceleration, e.g. by waves, the trajectories $E(t)$ are
not only non-monotonic but may  well be highly stochastic, differing
between electrons of the same initial $E_*$ (cf Section 3 for a
specific example). There is then no well defined deterministic yield
$\zeta(\epsilon,E_*)$ for electrons of initial $E=E_*$  and the
total yield has to be found numerically by evaluating expression
(\ref{zetadef}) for each electron and summing them, or using
statistical techniques \citep[e.g.][]{BianBrowning2008}.
\end{enumerate}

\section{Current Sheet Cascades (CSCs) as One Possible LRTTM Scenario}
{The LRTTM idea that local reacceleration of electrons inside the thick target HXR source can greatly increase their photon yield by prolonging their lifetimes to $\gg t_{coll}$ is a quite general one which might be realized for many different (re)acceleration mechanisms. The basic requirement is some source of strong electric fields distributed through the source and this might be achievable in a variety of ways - e.g.  \citet{lionello1998}, \citet{fletcher2008}. In this Section we focus on one possibility to illustrate the idea in some detail.

\begin{figure*}
\centering
  \includegraphics[width=17cm]{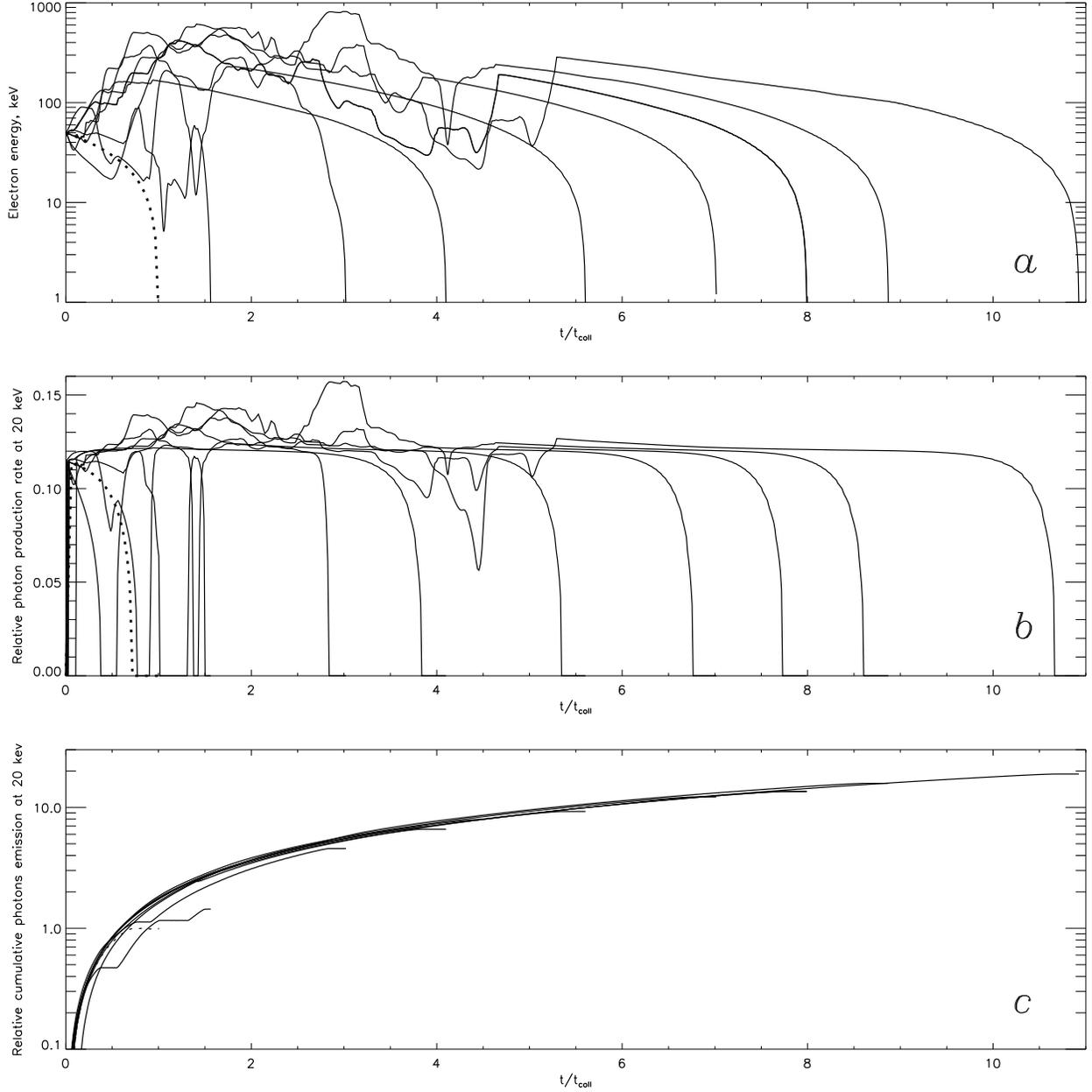}
   \caption{{\bf Panel a}. Examples of trajectories $E(E_*,t)$ for nine electrons re-accelerated in the chromosphere. and one (dotted curve) undergoing collisional losses only, after arriving from the corona with initial energy $E_* =50$ keV.
  {\bf Panel b}.  Relative photon production rate at 20 keV.
  {\bf Panel c}. Relative cumulative photon production $\int_0^t \dot\zeta(\epsilon,t)dt$ keV$^{-1}$ at $\epsilon =20$ keV for electrons shown in Panel a. Time is in units of the collision time for a 50 keV electron ($t_{coll}=0.0032$) s. The emission rates in b. and total emissions in c. have been divided by the total yield $\zeta_{CTTM}$ keV$^{-1}$ at 20 keV of a purely collisional electron starting at $E_*=50 keV$. In c., the $\zeta(t)$ curves that stop at some t have attained their asymptotic values there, the electron $E(t)$ having dropped below 20 keV thereafter}.
 \label{f2}
\end{figure*}

\subsection{Energy Release and Electron Acceleration in CSCs}
 The CTTM idea of separation of the acceleration and radiation volumes
  had its origins partly in the ideas that : energy is most easily
  stored in the corona \citep{Sweet1958};
acceleration is more efficient in a tenuous collisionless
volume \citep[e.g.][]{HamiltonPetrosian1992,Miller_etal1997}, while
bremsstrahlung gives most volumetric yield at high densities. In
such cases, magnetic energy release is assumed to be driven by
organized and continuous twist or shear of large scale magnetic
structures (isolated loops or arcades) - e.g. \citet{ForbesPriest1995}.
An alternative is distributed small scale release of energy
in a Current Sheet Cascade (CSC) \citep[][]{GalsgaardNordlund1997,Galsgaard2002},
resulting from the 3-D MHD response of a loop to a
random underlying photospheric driver. This gives a specific
physics-motivated example of the type of local re-acceleration
scenario discussed schematically in Section 2. After a few
Alfv\'{e}n times (secs), Lorentz forces create stresses along the
entire loop and form a hierarchy of reconnecting current sheets,
leading to plasma jets. These perturb the neighboring plasma and
eventually create a turbulent current sheet cascade (CSC) throughout
the volume from large scale current sheets (CSs) to numerous small
scale CSs in which energy is dissipated randomly everywhere. CSs
appear and disappear over short times but with an overall
quasi-steady turbulent state. This energy release and acceleration
process operates not only in the corona but also in the
chromosphere e.g. \citep[][]{Daughton_etal2008}.
Several papers have discussed particle acceleration in the CSC
electric fields $\cal E$ in such a dynamic environment \citep{AnastasiadisVlahos1994,Dmitruk_etal2003,ArznerVlahos2004,VlahosGeorgoulis2004,Vlahos_etal2004,Turkmani_etal2005,Turkmani_etal2006}, the last two of which simulated particle acceleration in the $\cal
E$ fields present in the coronal case of such models. For the
present paper we conducted similar calculations for already energetic
electrons injected into the dense
chromospheric part of the loop . Trajectories $E(t)$ of test
particles, with prescribed initial energy $E_*$ randomly injected in
space, were traced in a frozen "snapshot" of the MHD fields since
acceleration times ($\ll 1$ sec) are much shorter than the overall
MHD evolution timescale ($\sim$ seconds). Electrons gain or lose
energy stochastically as they travel along guide fields and pass
through the local CS $\cal E$-fields and lose it collisionally,
following complicated trajectories as they receive kicks of various
signs and strengths. Some test electrons encounter few or no CSs and
decay purely collisionally, rejoining the local thermal plasma.

Following \citet{Turkmani_etal2005,Turkmani_etal2006} we found that, in the corona, strong
acceleration of a substantial fraction of thermal electrons occurs for reasonable values of the resistivity $\eta$ so long as super-Dreicer ${\cal E}$ occur in some of the current sheets. In common with most acceleration modelling, it is hard to assess realistically how high this fraction is in a test particle approach involving scaled numerical resistivity and resolution. If the fraction becomes very high (as it has to in the CTTM to create enough HXRs) the validity of the MHD/test particle approach becomes questionable since the associated currents should be allowed to feed back on the MHD field equations. As we show below, in the LRTTM, the necessary $\cal F$ is reduced, which alleviates this issue. Our goal here is simply to show the  implications for HXR source requirements if extensive re-acceleration {\it does} occur.

After undergoing acceleration in the corona, electrons mainly precipitate into the
chromosphere where many of them, instead of rapid collisional decay, undergo re-acceleration in the chromospheric CS $\cal E$ fields. In this paper we therefore only
discuss what happens to these electrons once injected into the
chromosphere, namely  a substantial fraction of them survives at
high energies for many collision times, increasing the photon yield
$\zeta$ over the purely collisional CTTM value. This is a good example of the type of
LR scenario suggested schematically in Section 2 since the CTTM
distinction between acceleration and radiation regions disappears,
and (re-)acceleration occurs in the HXR source.

\subsection{CSC Simulations of Electron (Re)-Acceleration}

For the coronal part of the loop we adopted plasma values
$n=10^{10}$ cm$^{-3},T=10^6$K, $B=10^2$G and for the chromospheric
part $n=10^{13}$cm$^{-3},T=10^4$K, $B=10^3$G, though we recognise
that these vary in space (especially $n$ in the chromosphere). The depth of the chromosphere is taken to be
$7.5 \times 10^{3}$ km.
Here we neglect the effects of neutrals since at $T=10^4$K the chromospheric 
plasma is nearly full ionized. Future LRTTM modeling
with more realistic treatment of spatial structure should include the 
effect of neutrals in deeper cooler regions such as modifying the resistivity and
the collision rate. 
Collisions were treated using a modified form of $\dot E$ with the form $\dot
v/v\propto v^{-3}$,valid for $v\gg v_{th}$, multiplied by the factor $v^3/(v+v_{th})^3$ to avoid incorrect divergence near thermal speeds $v\simeq v_{th}$.

In the corona, after undergoing numerous CS $\cal E$-field
accelerations and decelerations, and collisions \citep{Turkmani_etal2005,Turkmani_etal2006}
many coronal electrons escaped
to the chromosphere, a situation geometrically similar to the
injection assumed in the CTTM with little collisonal HXR emission in
the tenuous corona. However, once in the chromosphere, as well as
collisions, electrons now undergo re-acceleration by the
CS $\cal E$ fields there. This greatly extends some of their
lifetimes beyond $t_{coll}$, increasing the mean photon yield
$\zeta$ and reducing the replenishment rate ${\cal F}_1$ hence the beam density
${\cal F}_1/Av_1$ needed over area $A$ to provide the HXR output
$J$, and so alleviating the problem of their large values in the CTTM.
In practice the corona accelerates and injects electrons with a spectrum
of $E_*$ - roughly a double power law distribution function \citep{Turkmani_etal2005,Turkmani_etal2006}.
Simulations of these spectral characteristics of injected electrons
and the resulting bremsstrahlung spectra arising from
the complex distribution of trajectories $E(E_*,t)$ in the thick target with re-acceleration (LRTTM)will be
the subject of future work. Here, to be able to compare simply the dynamics
and photon yields of electrons arriving in the chromosphere for  CTTM and LRTTM cases,
we limit our analysis to an ensemble of electrons all of the same $E_*$.

We have carried out such simulations of $E(t)$  for $10^3$ electrons
injected randomly with $E_*=50$ keV in a chromospheric CSC plasma
with ${\cal E},B$ fields from the MHD simulations discussed
in Section 3.1 and also for the purely collisonal case.
In some simulations, chromospheric CS $\cal E$ values were
high enough for some electrons in the tail of the local thermal
distribution to be accelerated but we do not consider these further here.
The simulation results depend on the electric field which can vary from one snapshot to another according to the dynamics of the turbulent loop. The main features
of our simulations are as follows :

\begin{enumerate}

\item {\it The accelerating electric field}: In the context of particle acceleration, electric fields are often compared with the Dreicer Field ${\cal E}_D$ (required for the force $e{\cal E}_D$ to overcome collisions for a thermal electron of $E\simeq kT$). Fast electrons arriving from the corona already have $E\gg kT$ in the chromosphere and the field required for re-acceleration to overcome collisions for them is smaller than ${\cal E}_D$ by a factor $kT/E$.

\item {\it The values of electric fields}: The electric fields are zero outside the current
sheets and found to take random values inside them. The average of this value
in the illustrative case used in this paper is $\cal E =$ $8.2 \times 10^{-4}$ statvolt/cm in the chromosphere  and its maximum value is ${\cal E}_{max} =$ $2 \times 10^{-2}$ statvolt/cm. The thickness of the current sheets vary between a minimum of
$0.5$ km and a maximum of $12.5$ km. 
However, comparison of cases in terms of 'average'
$\cal E$ values is not very meaningful. One could for example have
two cases with the same volume-averaged $\cal E$ in one of which
$\cal E$ nowhere approached $\cal E_D$ while in the other
$\cal E$ exceeded $\cal E_D$ in some local CSs.
Differences in the values of $\cal E $ affect the fraction of the injected electrons undergoing re-acceleration,
before being lost by escape or collisions. They also affect the maximum energies  electrons reach and their lifetimes.

\item {\it Direction of the fields}: Since the electrons encounter CS $\cal E$ fields in
quasi-random directions, the electrons move back and forth on guide
$B$ fields and their acceleration is stochastic with $\dot E$ undergoing
many changes of sign (cf. Section 2.3) as shown in Figure 2a. Though the electric
 fields are often high enough to accelerate or decelerate the electrons
inside the CSs, there is no global runaway because of the short
durations and quasi-random signs of these kicks. Scattering of the
electrons also helps enhance their lifetimes by keeping them in
the CSC region.

\item {\it Numerical Resistivity and Resolution}: The electric field considered here is the resistive electric field parallel component and its
value depends on the resistivity.  The average numerical resistivity used here
inside a chromospheric CS was taken to be $\eta = 3\times 10^{-13} s$
(roughly the Spitzer value) and, when combined  with the numerical resolution  used in our simulation, results in re-acceleration of electrons in the keV-MeV range most relevant to HXR burst production. Higher (anomalous) resistivities enhance the re-acceleration process. Increasing the resolution of the numerical 3D MHD experiment
leads to more and thinner fragmented CSs. This enhances the re-acceleration process
since the electrons undergo a higher number of smaller kicks, and
pass more often through electric field free zones in between.

\end{enumerate}

We found that, in the chromosphere, among the $10^3$ test electron CSC cases we ran, about 65 \% of injected
50 keV electrons underwent varying amounts of re-acceleration and lifetime enhancement well beyond the collision time $t_{coll}=3.2 \times 10^{-3}$ s.
Some examples of these re-accelerated electron trajectories $E(E_*,t)$ are shown
in Figure 2a for nine chromospherically re-accelerated electrons and for a CTTM electron.
It can be seen that electrons initially gain and lose energy in the CSs they
 randomly pass through, eventually entering electric field free zones where they escape or lose their energy to collisions. For our parameters, the increase in lifetimes of the injected electrons over $t_{coll}$ ranged from factors slightly higher than unity to around 20 $\times$
with an average over  all electrons of  about 5 $\times$.

\subsection{Numerical Results for Photon Yield $\zeta$ in the CSC LRTTM}

 As noted in Section 2.3, in the LRTTM scenario, electron trajectories
$E(E_*,t)$ starting from energy $E_*$ are not deterministic but
stochastic. Thus the only way to arrive at a measure of the photon
yield $\zeta(\epsilon,E_*)$ for a single test particle is to use its
individual equation of motion to compute $E(t)$ for use in time integration (\ref{zetadef}).
This is repeated for each of a sample of electrons initially of the same
energy $E_*$ but randomly located then
undergoing random kicks in the stochastic $\cal E$ fields. We want
to compare the mean $\zeta(\epsilon,E_*)$ of these with
the CTTM value $\zeta_{CTTM}(\epsilon,E_*)$. In the CTTM an
electron injected with $E_*\le\epsilon$ yields no photons of energy
$> \epsilon$ since $\dot E_{coll}<0$. This is not true in the
presence of re-acceleration since $E_{max}$ can exceed $E_*$. Some
criterion therefore has to be adopted for comparison of photon yields. Here we chose conservatively to compare the average photon
yields $\zeta(\epsilon,E_*)$ for the CTTM and CSC for electrons
launched inside the source from a specified initial high energy
$E_*> \epsilon$ namely 50 keV.

Based on Equation (\ref{zetadef}) we calculated for each test electron, using the accurate form of $Q_B$ from \citet{Haug1997}the rate $\dot \zeta$ of emission of photons per unit $\epsilon$ at 20 keV as a function of $t$
for each electron and also the cumulative number emitted up till $t$
per unit $\epsilon$ as a function of $t$, hence the total ($t\rightarrow \infty$) $\zeta$
value for each electron - see Figure 2 b, c.
In the unique CTTM case (Equation (\ref{zetadef})) for $E_*=50$ keV
$\zeta_{CTTM}(\epsilon,E_*)\approx 2.2\times 10^{-4}$ photons per keV at
$\epsilon= 20$ keV. The resulting mean increase in total photon yield per keV at 20 keV reached as high as a factor of around 20 with an average value around 10. These factors are higher than the increases in electron lifetimes mentioned above for the reason discussed in Section 2.3. For the given $E_*$ the resulting increase in yield $\zeta_{CSC}(\epsilon,E_*)$ compared to the collisional case increases with $\epsilon$.
 As we allow $\epsilon = 10, 20, 30, 40, 45, 49$ keV to approach $E_*$ the relative increase factors ${\zeta_{CSC}}/{\zeta_{CTTM}}$ in average yield
were about $5, 10, 15, 50, 100, 800$. This is because, in contrast with the monotonic CCTM fall of $E(t)$ from $E_*$, in the LRTTM many electrons of initial $E=E_*$ spend many times $t_{coll}$ at $E \gg E_*$, e.g.  50 - 800 keV in
the example shown in Figure 2. This produces many times more photons not only at 20 keV, as discussed above, but also at much higher $\epsilon$ whereas in the CTTM there is no photon yield above 50 keV. A proper comparison of LRTTM yield with CTTM is thus rather complicated and will require numerical simulations for many $E_*$ and $\epsilon$ values. But it is clear that the factors quoted above for the LRTTM photon yield enhancement are  conservative lower limits.

\section{Discussion and Conclusions}

 We have shown that substantial re-acceleration
 in the chromosphere of electrons accelerated
in and injected from the corona can greatly
reduce the density and number of fast electrons needed to produce a
HXR burst, and how this might occur in a CSC as one example.
In the LRTTM. as in the CTTM, most electron collisions are in
the chromosphere so the LRTTM also predicts HXR footpoints. Some of its other properties are, however, quite distinct
and need much more quantitative work beyond our outline ideas above for the
model to be evaluated and tested. Here we conclude by briefly
discussing some of the issues to be addressed.

\begin{enumerate}

\item {\it Fast Electron anisotropy}

In our CSC simulations we find that the electrons move more or less
equally up and down the loop axis ($<v_{z+}>\simeq <v_{Z-}>$) with $<v_\bot>/<v_\|>$ about 0.05 in
the chromosphere and 0.20 in the corona. Unlike the strong downward
beaming ($<v_{z+}>\gg<v_{z-}>$) in the basic CTTM \citep{brown1972}, this distribution is broadly
consistent with \citep{KontarBrown06mirror} albedo mirror diagnostic
'near isotropy' results from RHESSI spectra. The $v_\bot/v_\|\ll 1$
property of electrons in the CSC LRTTM may, however, still yield enough
$H_\alpha$ impact polarization to contribute to that
observed \citep{HenouxChambe1990,kasparova_etal2005} though other mechanisms
(e.g. fast proton impacts) may also contribute \citep{Henoux_etal1990}.

\item {\it HXR fine time structure and footpoint synchronism}

When fast electrons in the chromospheric HXR source originate by
injection from the corona, the HXR light curve should
reflect the coronal supply rate quite closely since even the LRTTM
extended fast electron lifetimes $\tau$ are short. So this scenario is consistent with HXR fine time
structure ($< 1$) s \citep{Kiplinger_etal1983},
footpoint synchronism findings \citep{Sakao_etal1996},
and energy-dependent time-of-flight delay
results \citep{Aschwanden2004}, provided that acceleration
in the coronal CSCs is coherent on short enough timescales. This
coherence should be on the coronal loop Alfv\'{e}n timescale
$\tau_A\simeq L/v_A \simeq 0.5L_9n_{10}^{1/2}/B_3$ so $B\simeq 500$
gauss suffices to make $\tau_A < 1$s.

\item {\it Interplanetary and HXR Flare Electron Fluxes
and Spectra}

In the CTTM the power law spectral index $\gamma$ of HXR emission
$J(\epsilon)$ is related to the spectral index $\delta_{thick}$ of
the electron injection rate ${\cal F}(E_o)$ by $\gamma =
\delta_{thick}-1$ and to the mean source electron flux $\overline
F(E)$ index $\delta_{thin}$ by $\gamma = \delta_{thin}+1$.
($\delta_{thick}-\delta_{thin}=2$ because the collisonal energy loss
cross section varies as $E^{-2}$). In the LRTTM situation
trajectories $E(t)$ are stochastic and average behaviour depends on
the specific CSC realisation so no such obvious simple relationship
exists. This complication also means that while integral
deconvolution of $J(\epsilon)$ \citep[e.g.][]{brown1971,Brown_etal2006}
to find the HXR source $\overline
nV\overline F(E)$ is still fully valid, inference of ${\cal
F}_*(E_*)$ is much more difficult because of the stochastic
character of the electron transport, in contrast with the simple CTTM
collisional case.

Using RHESSI and WIND data, \citep{Krucker_etal2007} studied the
relationship of electron spectra and numbers at the Sun to those
near the Earth above 50 keV. They find the indices and numbers to be
well correlated in all events involving free streaming from the Sun
but that the relationship of spectral indices does not
match the CTTM prediction \citep{Krucker_etal2007,Krucker_etal2009}.
Further, the numbers of electrons
in IP space and in Type III Bursts are smaller by a factor of order
500 \citep{Krucker_etal2007} than required for the HXR source in the
CTTM model. The
numbers required for microwave bursts are also often found to be considerably
less than for HXRs in the CTTM interpretation though this number is very sensitive
to assumed conditions (\citet{LeeGary2000} . For these to be consistent with the LRTTM reduction in electron
numbers would imply even more effective re-acceleration than in the illustrative example we gave here, such as due to anomalous resistivity.

\item {\it Impulsive flare heating}

Various aspects of impulsive flare heating data have been invoked in
support of the CTTM, including the Neupert Effect that flare soft XR
light curves correlate with the integral of HXR light
curves \citep{Neupert1968}. This is often attributed to CTTM collisional heating
of the SXR source plasma by HXR emitting fast electrons.
However, the observed relative time sequences of $EM(t), T(t)$  are hard to
reconcile with this in any obvious way even when filamented loop
structures are considered \citep{Veronig_etal2005,Stoiser_eta2008}.
In the LRTTM the total power delivered to fast electrons in
order to offset collisional losses is comparable to that in the CTTM
model and can likewise heat the impulsive flare atmosphere, though
the spatial distribution of that heating can be very different from
the CTTM case. In the proposed LRTTM scenario the coronally injected beam rate ${\cal F}_1$ is reduced considerably from the CTTM rate so beam heating of the corona is reduced from its CTTM value. However, for given HXR output, the total
beam power involved in the whole HXR source has to be at least as large as in the CTMM. Thus in the LRTTM more power goes into chromospheric heating as re-acceleration drives fast electrons against collisional
losses there. In addition, if the extended electron lifetimes result
in their penetrating deeper, beam heating may be effective to much
greater depths than in the CTTM. This might offer a solution to the
problem of heating white light flares by electron beams
\citep{Neidig1989,Fletcher_etal2007ApJ}.

\end{enumerate}

\acknowledgements We gratefully acknowledge financial support of this work by
a UK STFC Rolling Grant
(JCB, EPK, ALM) and Advanced Fellowship (EPK),
an EU Training Network (LV), a Royal Society Dorothy Hodgkin
Fellowship (RT), ISSI Bern (JCB,EPK,LV) and the Leverhulme Trust (EPK).
.

\bibliography{references}
\bibliographystyle{aa}

\end{document}